\documentclass[pra,twocolumn,showkeywords,longbibliography,superscriptaddress,10pt]{revtex4-1}
\usepackage{graphicx}
\usepackage{dcolumn}
\usepackage{color}
\usepackage{times}
\usepackage{bm}
\usepackage{amssymb}
\usepackage{amsmath}
\usepackage{epsfig}
\usepackage{epstopdf}
\usepackage{dsfont}
\usepackage{subfigure}
\usepackage{tikz}
\usepackage[colorlinks, citecolor=blue, linkcolor=blue,urlcolor=blue]{hyperref}
\usepackage[mathscr]{euscript}
\usepackage{orcidlink}
\usepackage{comment}
\usepackage{appendix}
\makeatletter
\renewcommand{\maketag@@@}[1]{\hbox{\m@th\normalsize\normalfont#1}}
\makeatother
\begin{document}
	\renewcommand{\thefootnote}{\fnsymbol {footnote}}
	
	\title{\textcolor{black}{Quantum steering as a probe of energy transfer in quantum batteries}}

	\author{Meng-Long Song~\orcidlink{0009-0002-9705-0241}}
	\affiliation{School of Physics, Anhui University, Hefei
		230601,  People's Republic of China}
	
	\author{Zan Cao}
	\affiliation{School of Physics, Anhui University, Hefei 230601, People's Republic of China}

    
	\author{Xue-Ke Song} 
	\affiliation{School of Physics, Anhui University, Hefei 230601,  People's Republic of China}
	
	\author{Liu Ye}
	\affiliation{School of Physics, Anhui University, Hefei 230601,  People's Republic of China}
	
	\author{Dong Wang~\orcidlink{0000-0002-0545-6205}} \email{dwang@ahu.edu.cn}
	\affiliation{School of Physics, Anhui University, Hefei 230601,  People's Republic of China}


	\date{\today}
	
	\begin{abstract}
This study investigates the role of EPR steering in characterizing the energy dynamics of quantum batteries (QBs) within \textcolor{black}{a charging system that features shared reservoirs. After optimizing parameter configurations to achieve high-energy systems, we observe across a variety of charging scenarios with low-dissipation regimes that steering serves as a vital resource: it is initially stored until the system reaches energy equilibrium, and then subsequently utilized to sustain the enhancement of energy storage. Furthermore, steering acts as a witness to battery population balance and a consumable that enhances extractable work. Additionally, we discuss the contribution of the steering potential to energy upon high-dissipation charging in details. These findings establish a novel indicator for monitoring QB energy variations, which will be beneficial to achieve the high-performance quantum batteries.}

	\end{abstract}
	
	\maketitle

	\section{Introduction}
	Quantum batteries (QBs) as an important topic of energy  in recent years has attracted a great deal of attention \cite{1,2}. Compared with classical batteries, QBs show their quantum advantages with an superextensive scaling maximum average power (i.e., QBs capture the most energy in the least amount of time) \cite{3,4,5}. On the one hand, given the potential of QBs as new energy devices, many excellent charging protocols have emerged. For examples, there are direct or charger-mediated charging protocols \cite{6,7,8,9,10,11,12,13,14,15,16,17}, as well as many-body charging protocols based on spin \cite{18,19,20,21,22}, Dicke model \cite{9,23,24,25}, Sachdev-Ye-Kitaev (SYK) model \cite{26,27}, and open charging protocols emphasizing dissipation \cite{28,29,30,31,32,33,34,35}. QBs are also implemented under a variety of experimental conditions, e.g., superconductors \cite{36,37}, quantum dots \cite{38}, etc \cite{39,40}. On the other hand, there have some interesting discussions about the nature of quantum batteries \cite{42,43,44,45, 46}. The results show that while quantum resources play a key role, it is difficult for a single resource (such as entanglement or coherence) to maintain a positive role all the time when a quantum battery is charging or extracting energy. The conclusion also extends to the relationships between the other quantum properties and the energy of the battery \cite{47,48}.  All of these indicate that it is necessary to discuss the role of quantum resources in specific charging protocols and settings, \textcolor{black}{which will enable us monitor and more accurately predict the energy transfer status of charging systems in complex charging protocols or scenarios, thereby achieving high-performance QBs.}
	
 Steering, one of the distinctive features of quantum mechanics, refers to the fact that a local measurement of one of the subsystems immediately disturbs the state of the distant other. After this concept was proposed by Schr\"{o}dinger \cite{49}, Wiseman \textit{et al}. \cite{50,51} clarified the definition of quantum steering (Einstein-Podolsky-Rosen (EPR) steering) in the local hidden state models. As a non-local quantum correlation, the steering is stronger than the quantum entanglement in the hierarchical relationship of quantum resource theories. EPR steering has been found not to be symmetric between the two subsystems \cite{52,53}, which indicate in some states, system \textit{A} can steer system \textit{B}, but \textit{B} does not necessarily steer \textit{A}. One of the fundamental problems is how to determine whether it is steerable in a certain quantum state. There have been much progress on this topic,  based on linearity \cite{54}, uncertainty relation \cite{55}, Clauser-Horne-Shimony-Holt–like (CHSH-like) inequalities \cite{56},  etc.

Therefore, we are curious as to what role the quantum steering plays during the quantum charging process, and whether it is available to probe energy transfer in the quantum battery protocol. Interestingly, the answer is found to be positive. Explicitly, 
we here concentrate on the \textcolor{black}{shared reservoir dissipation}  charging protocol, owing to its high average charging power of quantum batteries and the almost perfect useful work accumulation. \textcolor{black}{More importantly, it serves as a fundamental QB protocol that simulates the effects of different environments on the charging system; Consequently, the findings derived from this protocol will be applicable to a wider range of similar charging protocols.} Besides, we utilize a method for detecting the EPR steering based on the correlation matrix \cite{57,LiJuan}, which is stronger than the linear criterion. The performance of the charging system is investigated during energy transfer and the corresponding variation of steering. Several meaningful results are obtained as: (1) \textcolor{black}{High-energy charging system requires powerful pump and low-dissipation reservoirs, where charger is more sensitive to strong pumping drive, while battery energy storage demands a low-dissipation environment. As reservoir average particle number improves with rising temperature, the resulting increase in charging system energy overcomes the negative effects of system-pump detuning. Interestingly, detuning demonstrates its ability to enhance energy in high-temperature fermionic reservoirs.} (2) \textcolor{black}{Steering always seeks to maintain energy equilibrium within the charging system. In other words, the growth of the charging system's energy (including energy storage) invariably reserves a portion of steering until energy equilibrium is achieved, and then leverages the steering to further enhance itself.} (3) \textcolor{black}{Furthermore, steering's demand for energy balance extends to the population balance within the battery, and improvements in extractable work also require consuming existing steering.}

The structure of this paper is as follows. Section II briefly describes the charging scheme, \textcolor{black}{the definitions of the stored energy, ergotropy of QB,} and the steering criteria for the charging system. In Section III, \textcolor{black}{the energy of the steady-state system is investigated to identify optimal charging configurations.} In Section IV, we delve deeply into the role of steering in the energy of the charging system. Finally, our findings are summarized in Section V.
\\
\\

	\textcolor{black}{\section{General Theory}}

\begin{figure}
		
			\centering
			\subfigure{\includegraphics[width=8.5cm]{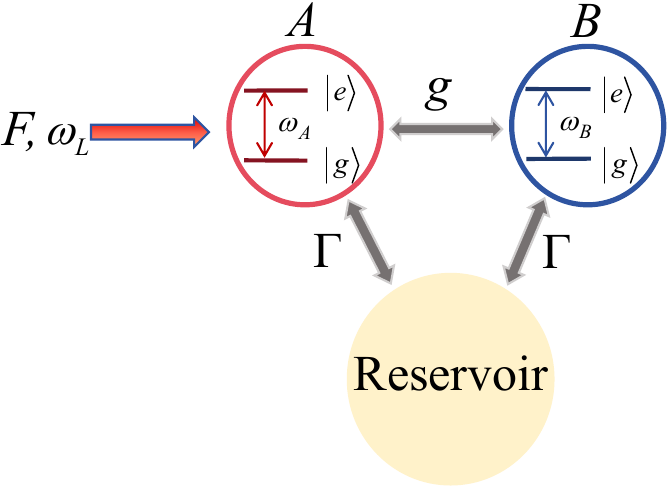}}
		\caption{As is shown, \textcolor{black}{a charger ($A$) coherently interacts with a battery ($B$) with a coupling rate $g$, where the frequencies of the charger and battery are $\omega_{A}$ and $\omega_{B}$, respectively. The charger and battery form a two-level system, where $\left| g \right\rangle $ represents the ground state and $\left| e \right\rangle $ the excited state. Both the charger and the battery simultaneously dissipate at a rate of $\Gamma$ into a common reservoir. And the charging system is powered by an external pump with a frequency of $\omega_{L}$ and an amplitude of $F$. The charging period is $t \in \left[ {0,\tau } \right]$, meaning that the battery reaches its maximum energy storage capacity at $t=\tau$.}}
		\label{fig1}
	\end{figure}

 \textcolor{black}{\subsection{Model}}
    
\textcolor{black}{The schematic diagram of the charging model is shown in Fig. \ref{fig1}. As illustrated, the charger is directly connected to the battery, and the charging system dissipates energy into an unspecified reservoir. Specifically, we consider bosonic or fermionic reservoirs at different temperatures, where the reservoir then serves as a new energy source for the battery alongside the charger. In the rotating frame, the Hamiltonian of the charging system $H_s$ can be written as ($\hbar=1 $ and the Boltzmann constant $k_B=1$):
\begin{align}
		H_s = \sum\limits_{i = A,B} {\frac{{{\Delta _i}\sigma _i^z}}{2}}  + g\left( {\sigma _A^ + \sigma _B^ -  + \sigma _A^ - \sigma _B^ + } \right) + F\left( {\sigma _A^ +  + \sigma _A^ - } \right).
		\label{Eq.H}
	\end{align}
where ${\Delta _i} = {\omega _i} - {\omega _L},(i=A,B)$ is the detuning of the charging system with a transition frequency of $\omega_i$ to the external pump with frequency $\omega_L$, we set the charger and battery to the same frequency for simplicity, i.e., ${\omega _A} = {\omega _B} = {\omega _0}$, thus ${\Delta _A} = {\Delta _B} = \Delta $. Moreover, $\sigma _i^{x,y,z}$ are the Pauli operators, $\sigma _i^{+,-}$ are the upper and down operators. The interaction between the charger and the battery is established through a coherent coupling with a rate of $g$, and $F$ is the amplitude of pump which is utilized to provide energy to the charger.}

\textcolor{black}{Assume that the shared reservoir is Markovian, thus the Lindblad quantum master equation is employed to describe the dynamics of the charging system as \cite{14,58}
\begin{align}
		{{\dot \rho }_{AB}} &= -i\left[ {H_s},{{\rho _{AB}}} \right] \\ \nonumber
        &+ \Gamma \left\{ {N\left( T \right)\sum\limits_{j = \sigma _{A,B}^ - } {{L_j}\left[ {{\rho _{AB}}} \right]}  + n\left( T \right)\sum\limits_{j = \sigma _{A,B}^ + } {{L_j}\left[ {{\rho _{AB}}} \right]} } \right\}
	\end{align}
where the dissipative superoperator ${L_j}\left[ \rho  \right] = j\rho {j^\dag } - \left\{ {{j^\dag }j,\rho } \right\}/2$, $\Gamma$ is the dissipative rate and ${n\left( T \right)}$ is the average particle number of the reservoir with frequency $\omega_k$ and temperature $T$. Specifically, $n\left( T \right) = 1/\left[ {\exp \left( {{\omega _k}/T} \right) - 1} \right] = {n_b}$ and $N\left( T \right) = 1 + {n_b}$ in
the bosonic reservoir, while $n\left( T \right) = 1/\left\{ {\exp \left[ {\left( {{\omega _k} - \mu } \right)/T} \right] + 1} \right\} = {n_f}$ and $N\left( T \right) = 1 - {n_f}$ for the fermionic reservoir, where $\mu$ denotes the chemical potential of the reservoir.}

\textcolor{black}{\subsection{Figures of merit}}

To evaluate battery performance, two fundamental metrics can be referred as: the energy stored $E_B$ within the battery and the extractable work ergotropy $W_B$ it contains. Set the battery to its initial ground state,  so that the stored energy can be read as
\begin{align}
		{E_B}\left( t \right) = {\rm{Tr}}\left[ {{H_B}{\rho _B}\left( t \right)} \right],
		\label{Eq.internal_energy}
 \end{align}
where $H_B = \omega_B \sigma_A + \sigma_B$ and $\rho_{AB}(t) = \operatorname{Tr}_A [\rho_{AB}(t)]$ is the reduced density matrix of $\rho_{AB}(t)$. Specifically, ergotropy is defined as the maximum energy that can be extracted by QBs in cyclic unitary operations \cite{65,66,67}: $W_B(t) = E_B(t) - \operatorname{Tr}[H_B \rho_p]$, where $\rho_p$ is the passive state of $\rho_B(t)$ without which energy can be extracted by cyclic unitary operations. In spectral decomposition, the Hamiltonian $H$ and density matrix $\rho$ are written as $H = \sum\nolimits_m {{e_m}\left| {{e_m}} \right\rangle \left\langle {{e_m}} \right|} $ and $\rho  = \sum\nolimits_m {{\lambda _m}\left| {{\lambda _m}} \right\rangle \left\langle {{\lambda _m}} \right|} $, respectively, where
$e_m$ and $\lambda_m$ are the eigenvalues of $H$ and $\rho$ with corresponding eigenstates ${\left| {{e_m}} \right\rangle }$ and ${\left| {{\lambda_m}} \right\rangle }$. The passive state can be derived as ${\rho _p} = \sum\nolimits_m {{\lambda _m}\left| {{e_m}} \right\rangle \left\langle {{e_m}} \right|} $ where ${\lambda _m} \geqslant {\lambda _{m + 1}}$ and ${e_m} \leqslant {e_{m + 1}}$. Then the ergotropy can be represented as 
   \begin{align}
		{W_B}\left( t \right) = {E_B}\left( t \right) - \sum\nolimits_m {{e_m}{\lambda _m}}.
		\label{Eq.ergotropy}
	\end{align}
To be specific, ${E_B}\left( t \right) = {\omega _0}\left[ {{\rho _{11}}\left( t \right) + {\rho _{33}}\left( t \right)} \right]$ and ${W_B}\left( t \right) = {\omega _0}\left\{ {{{\left\{ {4{{\left| {{\rho _{12}}\left( t \right) + {\rho _{34}}\left( t \right)} \right|}^2} + {\chi ^2}} \right\}}^{1/2}} + \chi } \right\}/2$ for the charging system under consideration, where ${\rho _{ij}}\left( t \right)$ is the $i$th row and $j$th column matrix entry of ${{\rho _{AB}}\left( t \right)}$ with $\chi  = 2\left[ {{\rho _{11}}\left( t \right) + {\rho _{33}}\left( t \right)} \right] - 1$.

Generally, the internal energy of the charger $A$ can be described as
\begin{align}
		{E_A}\left( t \right) = {\rm{Tr}}\left[ {{H_A}{\rho _A}\left( t \right)} \right],
		\label{Eq.charger}
\end{align}
where ${H_A} = \omega_0 \sigma _A^ + \sigma _A^ - $ and ${\rho _A}\left( t \right) = {\rm{Tr}}_B \left[ {{\rho _{AB}}\left( t \right)} \right]$. Further, ${E_A}\left( t \right) = {\omega _0}\left[ {{\rho _{11}}\left( t \right) + {\rho _{22}}\left( t \right)} \right]$ for the charging system.

In addition, we refer to the practice of using local orthogonal observables  \cite{63} (LOOs) to detect EPR steering. Specifically, for a bipartite state $\rho $ on ${\wp ^{{d_1}}} \otimes {\wp^{{d_2}}}$, if ${G_{set}} = \left\{ {{G_i}:i = 1,2, \ldots ,d_1^2} \right\}$ and ${\Lambda _{set}} = \left\{ {{\Lambda _j}:j = 1,2, \ldots ,d_2^2} \right\}$ be any LOOs set of $\wp^{{d_1}}$ and $\wp^{{d_2}}$, respectively. And the correlation matrix formed by LOOs is $C\left( {{G_{set}},{\Lambda _{set}}\left| \rho  \right.} \right) = \left\{ {{\rm tr}\left[ {\left( {{G_i} \otimes {\Lambda _j}} \right)\left( {\rho  - {\rho _A} \otimes {\rho _B}} \right)} \right]} \right\}$. In this case, the EPR steering operation criterion can be written as \cite{57}: if the state $\rho$ on ${\wp ^{{d_1}}} \otimes {\wp^{{d_2}}}$ is \textit{unsteerable} from $A$ to $B$, then 
\begin{align}
\left\| C\left( {G}_{\text{set}}, \Lambda_{\text{set}} \middle| \rho \right) \right\|_{\mathrm{tr}} 
    \leq \sqrt{ \left[ d_1 - \operatorname{tr}\left( \rho_A^2 \right) \right] \left[ 1 - \operatorname{tr}\left( \rho_B^2 \right) \right] },\\
		\label{Eq.unsteerable}
\end{align}
where ${\left\| C \right\|_{\rm{tr}}}$ is the trace norm of $C$. In other words, its converse-negative proposition is: if Eq. (\ref {Eq.unsteerable}) is violated or the difference ${S_{A\to B}} = {\left\| C \right\|_{\rm{tr}}} -  {\sqrt {\left[ {{d_1} - {\rm{tr}}\left( {\rho _A^2} \right)} \right]\left[ 1 - {{\rm{tr}}\left( {\rho _B^2} \right)} \right]}}  >0 $, it means that $A$ can steer $B$. Similarly, when the difference $S_{B\to A}>0$, it also indicates that $B$ can steer $A$. For convenience, we call $S_{A\to B, (B\to A)}$ as the steering function and employ its magnitude to quantify the steering potential of the charger or battery, that is, the increase of $S_{A\to B, (B\to A)}$ means the improvement of the steering potential, and the steering will not appear until the steering potential is activated $S_{A\to B, (B\to A)}>0$. Further, we consider LOOs as the set $\left\{ {I,{\sigma ^x},{\sigma ^y},{\sigma ^z}} \right\}/\sqrt 2 $ to detect the steerability of ${\rho }\left( t \right) = \left| {\psi \left( t \right)} \right\rangle \left\langle {\psi \left( t \right)} \right|$, \textcolor{black}{where $I$ is the identity matrix.}

\textcolor{black}{\section{Battery performance}}

\textcolor{black}{Before investigating the contribution of steering to energy, a preparatory step is required: ensuring the charging system possesses sufficient energy. This guarantees the battery has the opportunity to reach its energy ceiling while enhancing the system's resistance to dissipation. Therefore, we focus on pursuing optimal energy parameter settings in this section. Specifically, we examine the system's steady-state energy to mitigate the interference of initial conditions on energy.}

\textcolor{black}{By setting steady-state conditions ${{\dot \rho }_{AB}} = 0$, we can determine the state ${\rho _{AB}}\left( \infty  \right)\left| {_{\Delta  = T = 0}} \right.$ of the charging system, thereby obtaining an analytical expression for the system's energy:
\begin{align}
  \begin{gathered}
  \frac{{{E_A}\left( \infty  \right)}}{{{\omega _0}}} = 4{k^2}\left[ \begin{gathered}
  32{k^4}\left( {1 + {l^2}} \right) + {l^4}\left( {4 + 9{l^2}} \right) \hfill \\
   + 4{k^2}\left( {2 + {l^2}} \right)\left( {4 + 9{l^2}} \right) \hfill \\ 
\end{gathered}  \right]/\alpha , \hfill \\
   \frac{{{E_B}\left( \infty  \right)}}{{{\omega _0}}} = 16{k^2}\left[ {8\left( {{k^2} + {k^4}} \right) + 2\left( {2 + 9{k^2}} \right){l^2} + 9{l^4}} \right]/\alpha , \hfill \\ 
\end{gathered} 
		\label{Eq.steady_energy}
\end{align}
where $\alpha  = 256{k^6}\left( {1 + {l^2}} \right) + {l^2}{\left( {4 + {l^2}} \right)^2}\left( {4 + 9{l^2}} \right) + 4{k^2}{l^2}\left( {4 + 3{l^2}} \right)\left( {4 + 9{l^2}} \right) + 64{k^4}\left( {4 + 11{l^2} + 5{l^4}} \right)$ with $k=F/g$, $l=\Gamma/g$.}

\begin{figure}
\begin{minipage}{0.5\textwidth}

		\centering
		\subfigure{\includegraphics[width=4.3cm]{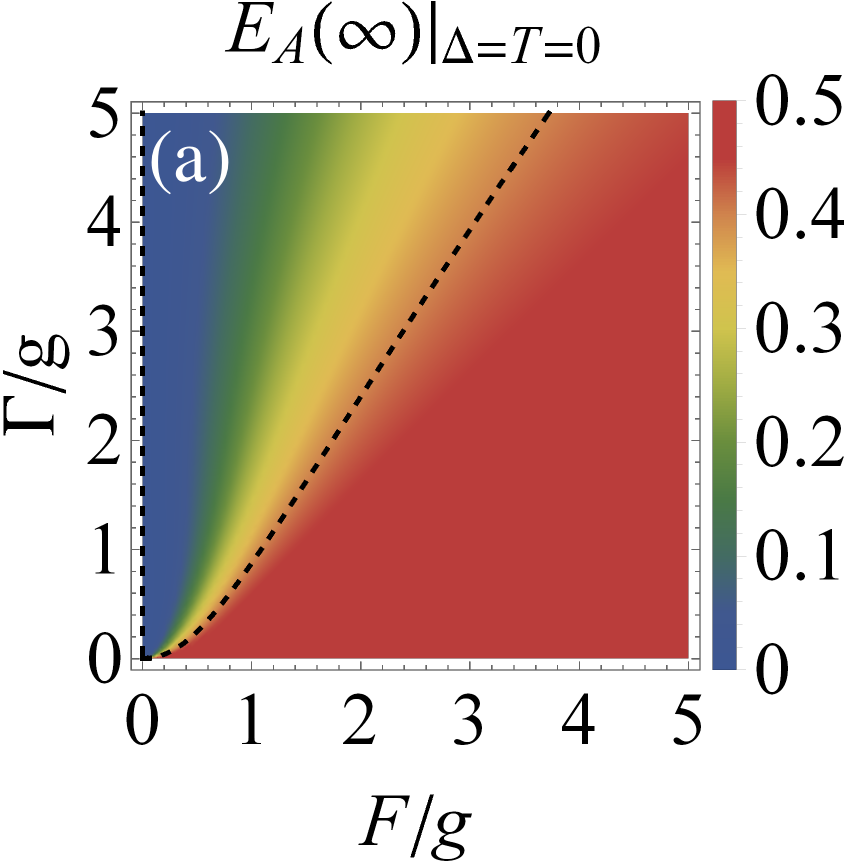}}
		\hspace{0.1cm}
		\subfigure{\includegraphics[width=4.3cm]{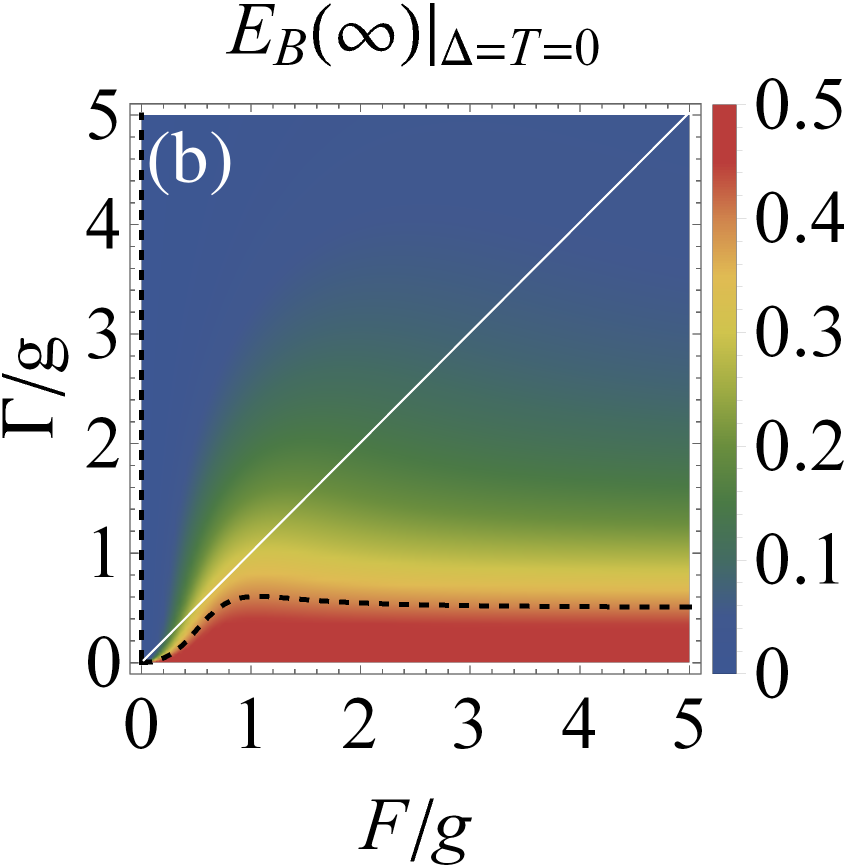}}
       
		\caption{\textcolor{black}{The steady-state energy ${E_{A,B}}\left( \infty  \right)$ (units of $\omega_0$) of the system as a function of pump $F$, dissipation rate $\Gamma$, and interaction strength $g$. Graph (a): the charger's energy ${E_{A}}\left( \infty  \right)$, and Graph (b): the battery's energy ${E_{B}}\left( \infty  \right)$. Here we have set up a resonant pump and a reservoir at zero-temperature, i.e., $\Delta  = T = 0$. The black dashed lines represent contour lines, while the white line in (b) depict energy pathway resulting from adjusting the coupling strength $g$ when the $F=\Gamma$. Specifically, the continuously increasing coupling strength will lead to the decrease of $\Gamma/g$ and $F/g$.}}
		\label{fig2}
        \end{minipage}
	\end{figure}  
\begin{figure}
		\begin{minipage}{0.5\textwidth}
			\centering
		\subfigure{\includegraphics[height=4.3cm]{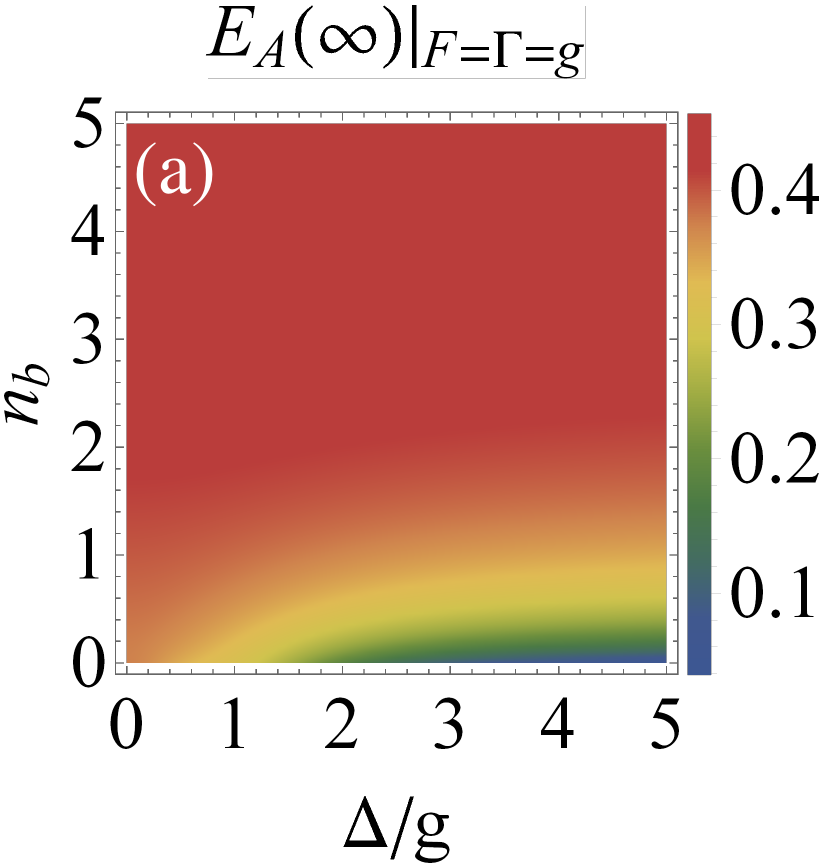}}
		\hspace{0.01cm}
		\subfigure{\includegraphics[height=4.3cm]{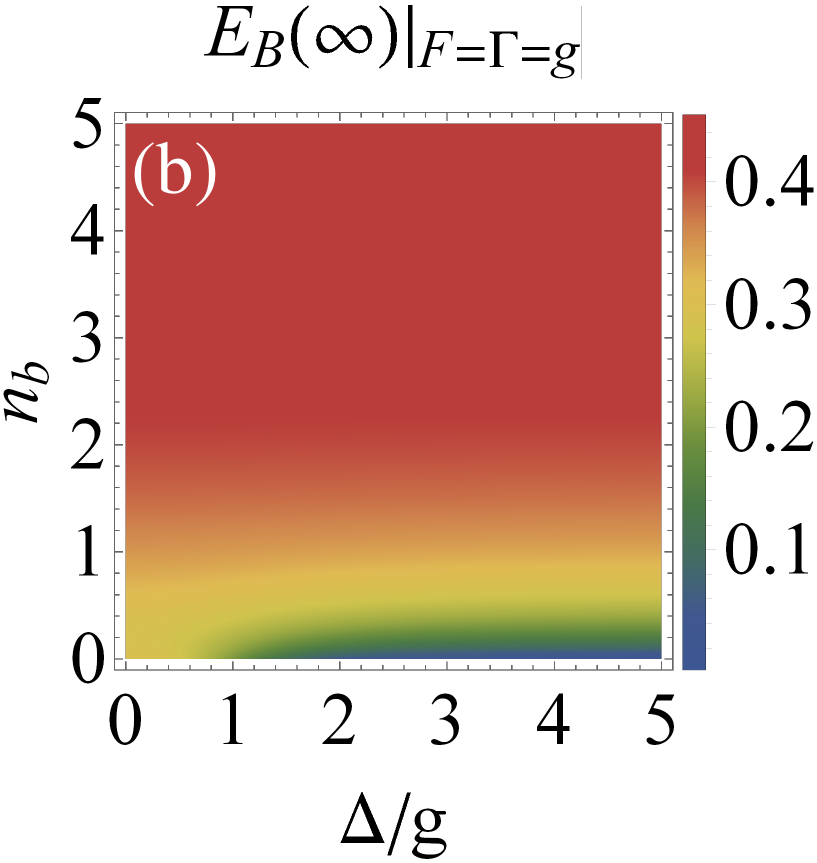}}
		\\
		\subfigure{\includegraphics[width=4.3cm]{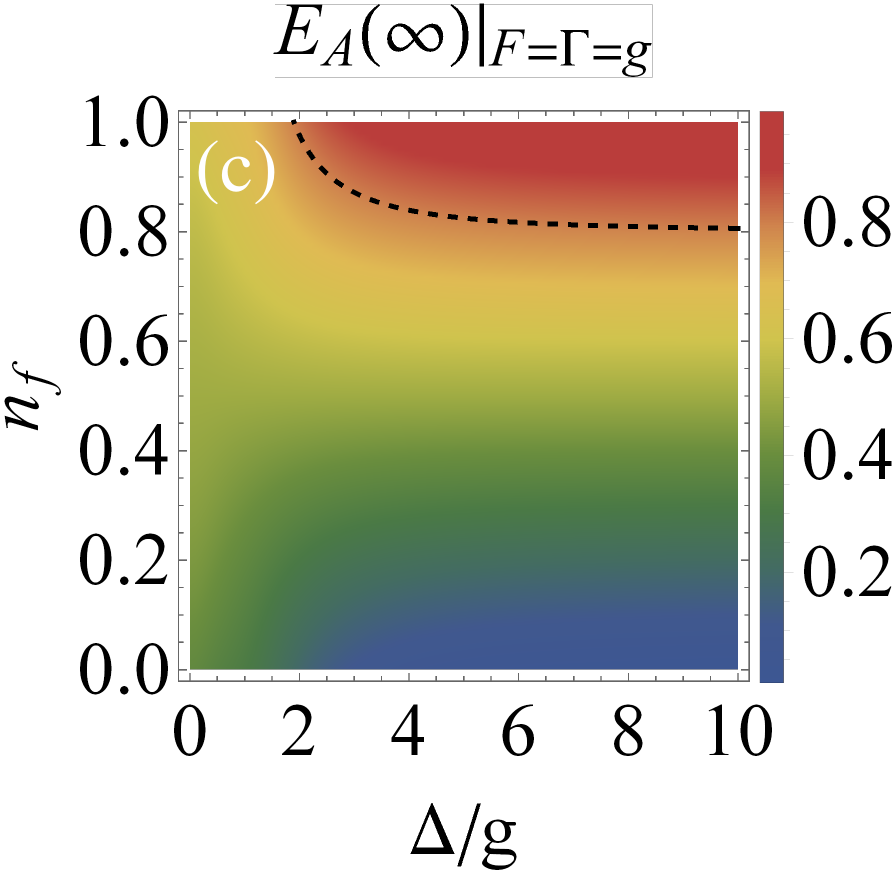}}
		\hspace{0.01cm}
		\subfigure{\includegraphics[width=4.3cm]{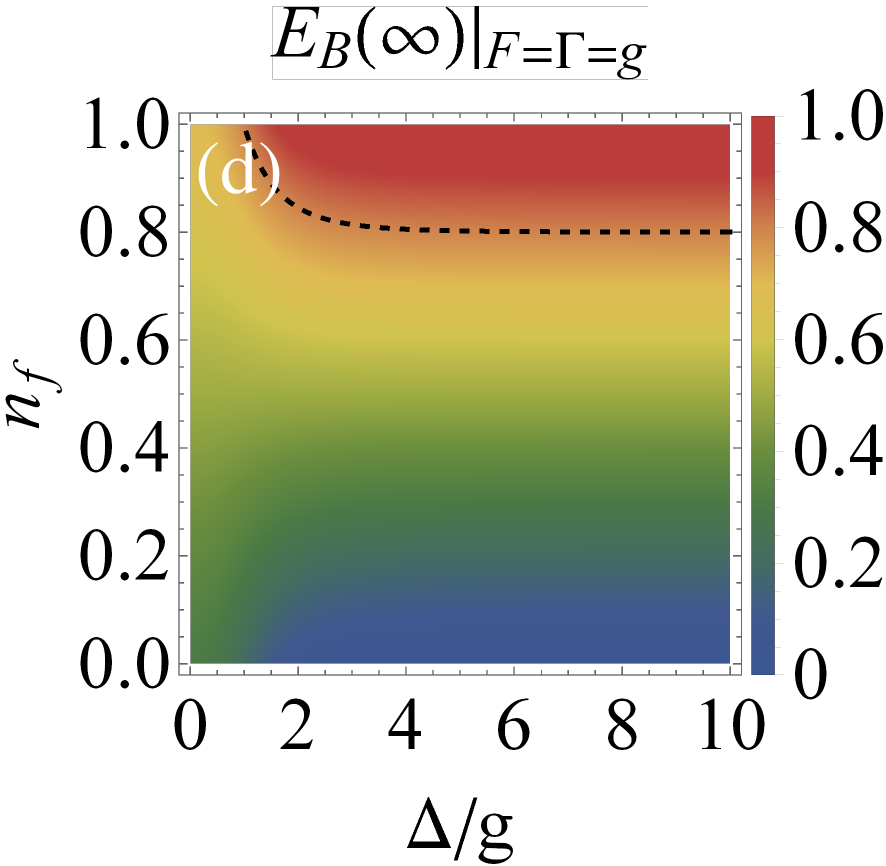}}
      
		\end{minipage}\hfill
		\caption {\textcolor{black}{The system's energy ${E_{A,B}}\left( \infty  \right)$ (units of $\omega_0$) as a function of the average particle number $n_{b,f}$ and the detuning $\Delta$. Here we have considered different types of non-zero-temperature reservoirs and the effect of detuning on energy. Here we have examined different types of non-zero-temperature reservoirs and the impact of detuning on energy. Since temperature elevation directly improves the average particle number $n(T)$ in the reservoir, we intuitively demonstrate how variations in $n(T)$ affect energy. Note that symmetric driving, dissipation, and coherent coupling are set here (i.e, $F=\Gamma=g$) with charger energy plotted in (a) and (c), and energy storage depicted in (b) and (d).}}
		\label{fig3}
	\end{figure}

\textcolor{black}{Fig. \ref{fig2} depicts the energy of a steady-state system as functions of $\Gamma/g$ and $F/g$. It notes that the disappearance of the average particle number (i.e., $n(T)=0$) caused by zero temperature prevents the reservoir from functioning as an energy source for the battery. As the sole energy input, the pump simultaneously supports both the charger and the battery. Strong driving serves as the cornerstone for preventing energy dissipation within the system and promoting energy accumulation in the battery. Furthermore, adjusting the coherent coupling $g$ alone will yield complex effects on energy. For instance, gradually increasing coherent coupling when the driving strength and dissipation rate are symmetric (i.e., $F=\Gamma$) leads to a slow accumulation followed by a rapid loss of stored energy in the battery. This behavior is illustrated by the battery energy changes corresponding to the white line in Fig. \ref{fig2}(b), suggesting that stronger coupling is not necessarily more favorable for energy storage.}

In addition, while strong drive and weak dissipation are recognized as key factors for enhancing charger energy and energy storage, these two settings yield different effects on energy improvement. In Fig. \ref{fig2}(a), the contour lines are more skewed toward the $\Gamma/g$, indicating that the charger's energy is more sensitive to $F/g$. That is, enhancing the drive yields a greater energy gain relative to reducing dissipation. In contrast, as shown in Fig. \ref{fig2}(b), energy storage exhibits high sensitivity to $\Gamma/g$. On one hand, the energy pumped into the charger will scarcely reach the battery for storage, if low-dissipation charging cannot be achieved. On the other hand, further driving enhancement does not improve energy storage  under low-dissipation regime. 

\textcolor{black}{Figs. \ref{fig3}(a)-(b) illustrate the energy variation of the charging system in a non-zero-temperature boson reservoir. Notably, the disruptive effect of detuning on energy is confined to low-temperature scenarios. As temperature increases (with a rise in the $n_b$), the reservoir functions as an energy source, supplying the charging system and thereby counteracting the negative effects of detuning. At this point, detuning can no longer stimulate energy variation. Meanwhile, Figs. \ref{fig3}(c)-(d) show the behavior of the system's energy in a non-zero-temperature fermionic reservoir. The results indicate that the effect of detuning on energy depends on the reservoir temperature: greater detuning degrades energy at low temperatures, whereas at high temperatures, greater detuning enables the charging system to achieve its maximum energy. To be more specific,  a critical temperature transition occurs where the $n_f=0.5$. Reservoirs  with $n_f<0.5$ are classified as low-temperature reservoirs, while $n_f>0.5$ are designated as high-temperature reservoirs.}

\textcolor{black}{ \section{The Effect of EPR steering on Energy}}

\textcolor{black}{After obtaining the parameters required for optimal energy, this section will turn to investigate the contribution of steering between the charger and battery to system energy during the charging system evolution. By simulating various charging scenarios, we  reveal the role of steering in energy improvement.}

\begin{figure}
		\begin{minipage}{0.5\textwidth}
			\centering
		\subfigure{\includegraphics[height=5cm]{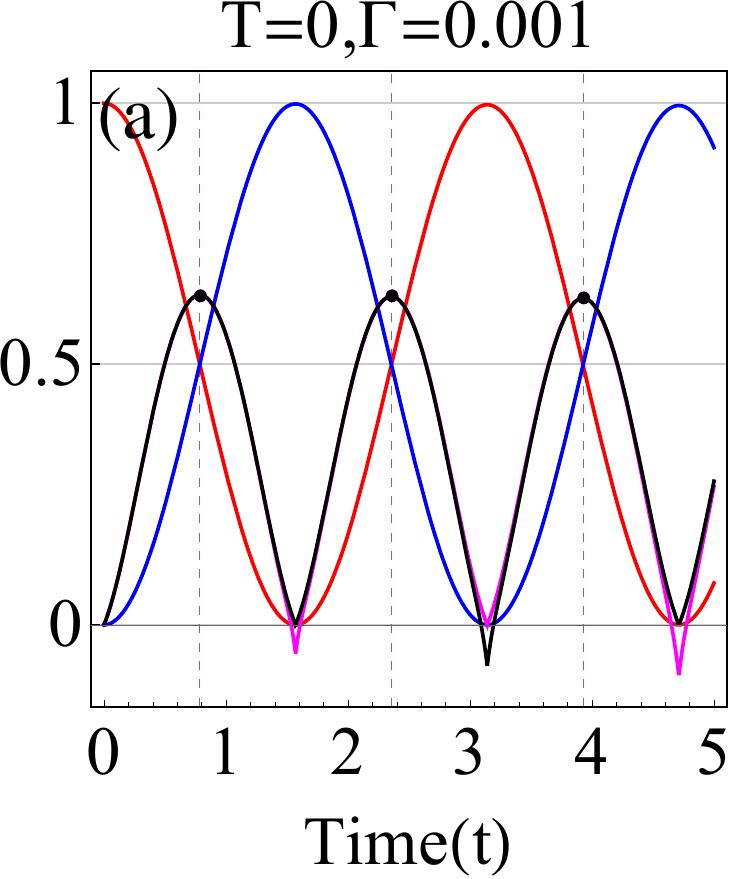}}
		\hspace{0.1cm}
		\subfigure{\includegraphics[height=5cm]{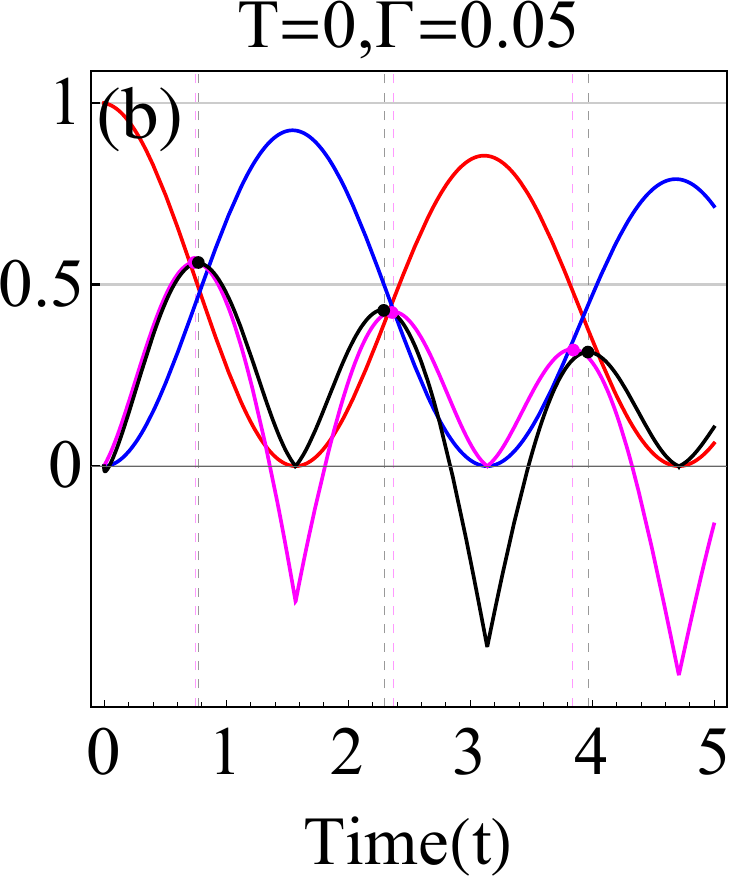}}
		\\
		\subfigure{\includegraphics[height=5cm]{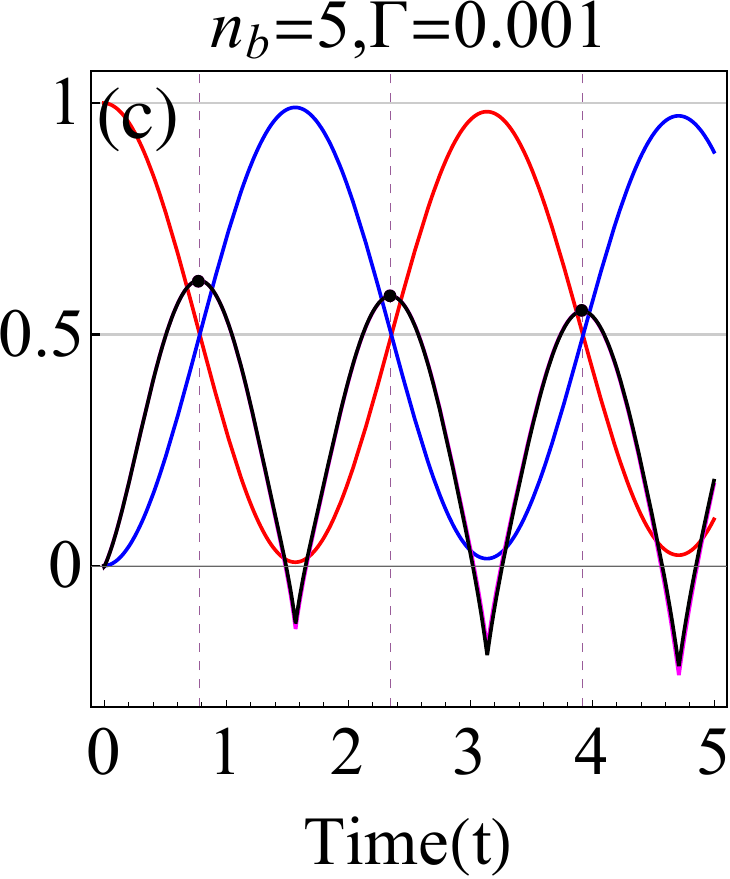}}
		\hspace{0.1cm}
		\subfigure{\includegraphics[height=5cm]{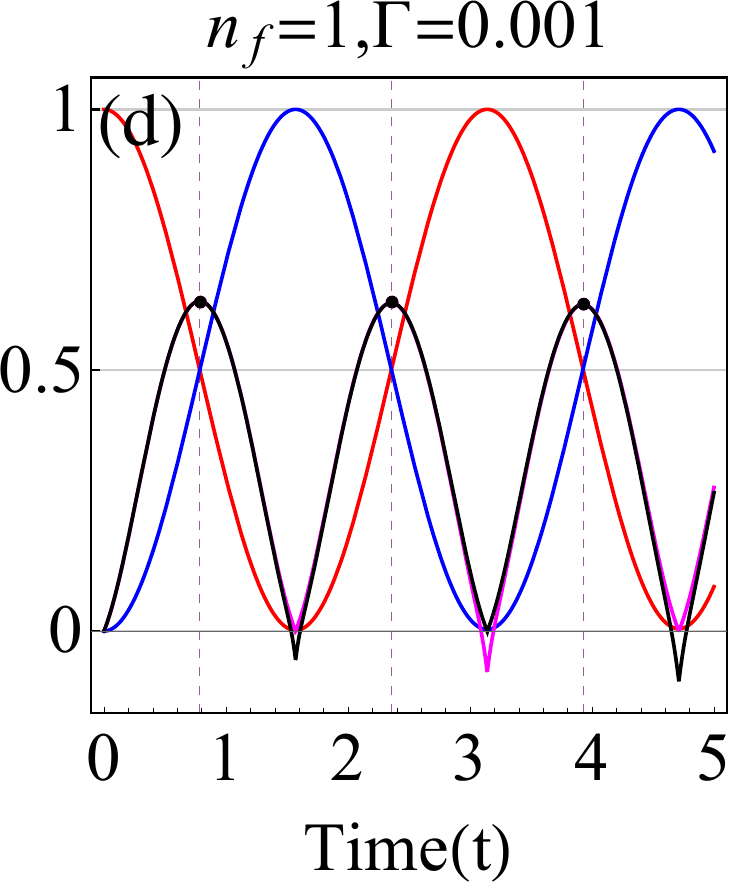}}
          \\
        \subfigure{\includegraphics[width=5cm]{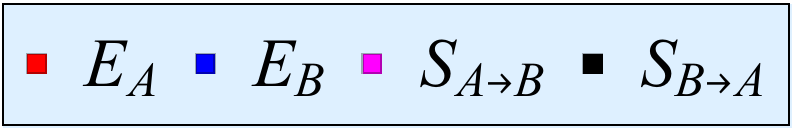}}
		\end{minipage}\hfill
		\caption {\textcolor{black}{The system's energy ${E_{A,B}}/ \omega_0$ and steering function $S_{A\to B,(B\to A)}$ as functions of time. Here, (a) and (b) account for resonance-driven processes and zero-temperature reservoirs, while (b) emphasizes stronger system dissipation. (c) and (d) set corresponding parameters for reservoirs at different temperatures to achieve maximum system energy. Since (a)–(d) all belong to the no-pumping charging scenario ($F=\Delta=0, g=1$), we selected a fully charged initial charger and a completely depleted battery, i.e., $\left| {\psi  \left( 0 \right)} \right\rangle  = \left| {eg} \right\rangle  $. Note that the magenta dots and black dots represent the maxima of steering functions $S_{A\to B}$ and $S_{B\to A}$, respectively, while the corresponding dashed lines indicate the time points at which these maxima occur.}}
		\label{fig4}
	\end{figure}

 \textcolor{black}{\subsection{Pumpless Charging ($F=0$)}}

\begin{figure*}
		\begin{minipage}{1.0\textwidth}
			\centering
		\subfigure{\includegraphics[width=5.3cm]{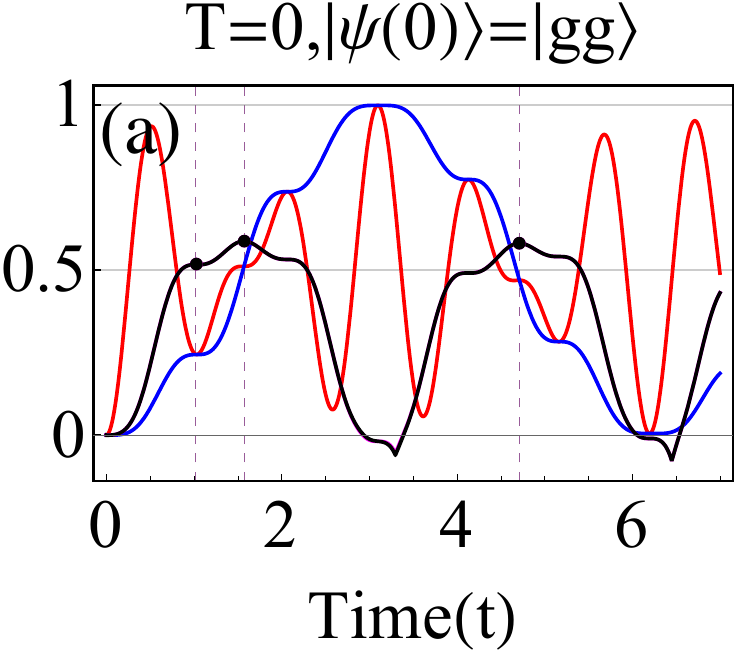}}
		\hspace{0.01cm}
		\subfigure{\includegraphics[width=5.3cm]{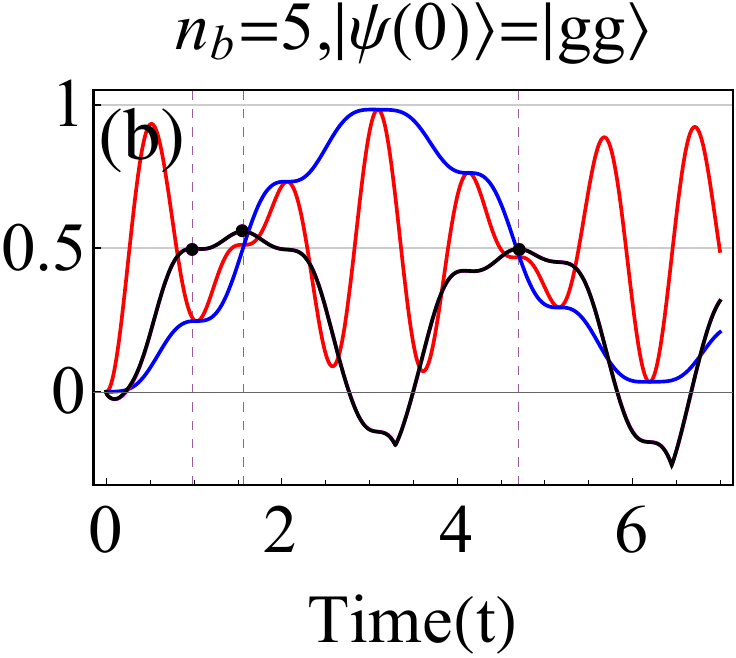}}
        \hspace{0.01cm}
        \subfigure{\includegraphics[width=5.3cm]{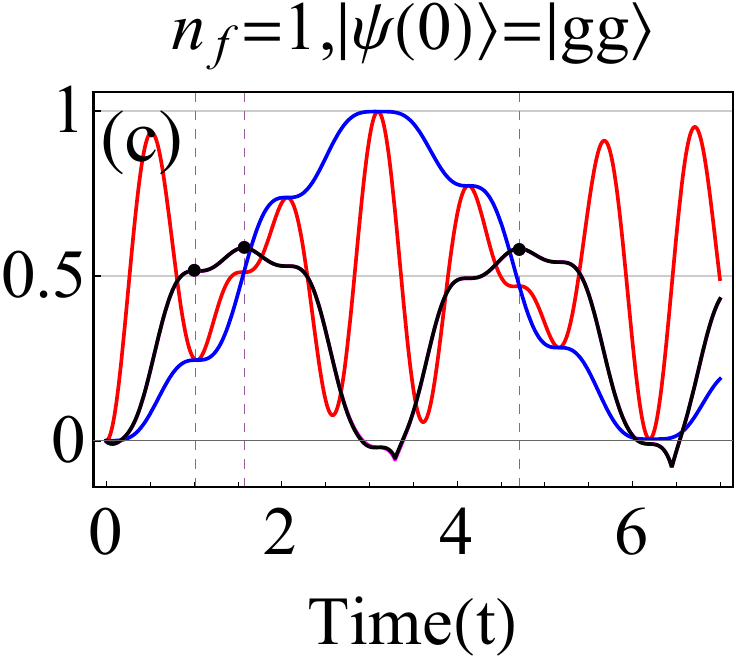}}
		\\
		\subfigure{\includegraphics[width=5.3cm]{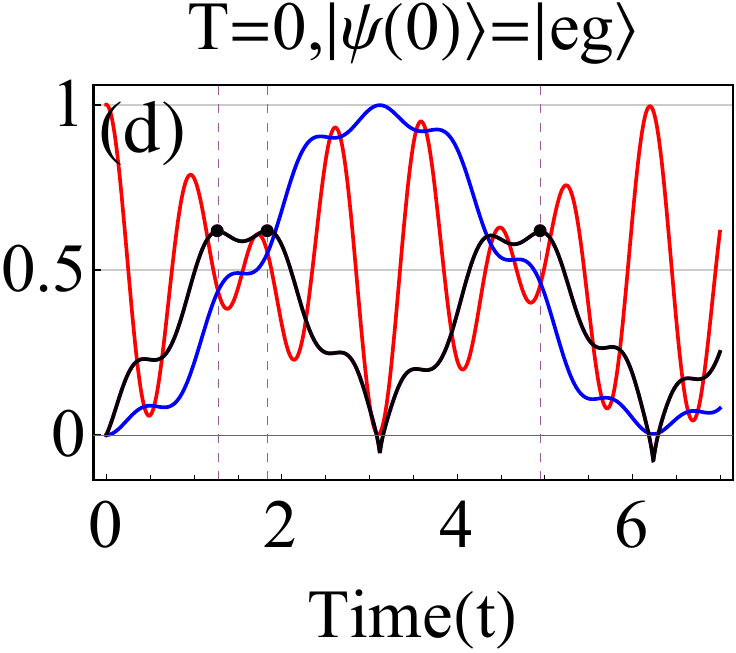}}
		\hspace{0.01cm}
		\subfigure{\includegraphics[width=5.3cm]{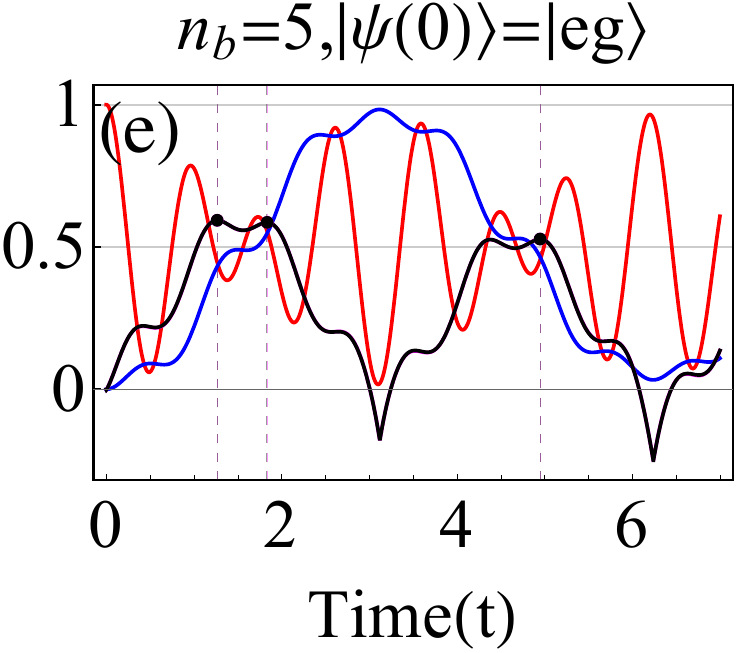}}
        \hspace{0.01cm}
        \subfigure{\includegraphics[width=5.3cm]{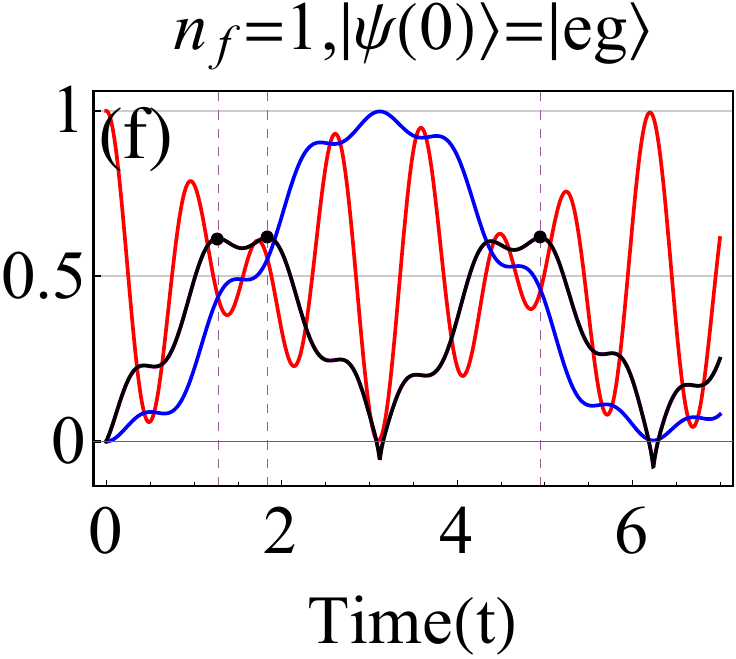}}
        \\
         \subfigure{\includegraphics[width=6cm]{457_lengend.pdf}}
		\end{minipage}\hfill
		\caption{ \textcolor{black}{The time evolution of the energy $E_{A,B}/\omega_0$ and steering function $S_{A\to B}$ and $S_{B\to A}$ are plotted here. We considered resonance-driven charging scenarios ($F=3,\Gamma=0.001,\Delta=0,g=1$), including different initial conditions and the influence of reservoir temperature on energy. Note that the magenta dots and black dots represent the maxima of steering functions $S_{A\to B}$ and $S_{B\to A}$, respectively, while the corresponding dashed lines indicate the time points at which these maxima occur.}}
		\label{fig5}
	\end{figure*}
    
\textcolor{black}{In this subsection, we examine the energy transfer dynamics of the system in the absence of the pump. To ensure sufficient internal energy within the system, we select a charger initially in an excited state and also incorporate the results from the parameter optimization conducted in the previous section.}

\textcolor{black}{Figs. \ref{fig4}(a)-(b) illustrate the evolution of charging systems in zero-temperature reservoir. Graph (a) highlights the low-dissipation regime, where the charging system operates almost decoupled from the reservoir like a closed system, with energy flowing solely between the charger and the battery. Moreover, the charger or battery can always steer the other party, except in one special case: when energy is completely transferred to one side. On the one hand, observing this special case reveals that energy is closely related to steerability, that is, low-energy subsystems often lose their ability to steer the other party. For example, $S_{A\to B}<0$ for $E_A=0$ or $S_{B\to A}<0$ for $E_B=0$.  On the other hand, we observe that the maximum steering occurs only when the charger and battery possess identical energy. Once either deviates from this energy equilibrium, steering is suppressed regardless of which direction the energy shift occurs. This indicates that initial battery charging requires accumulating steering, while higher energy storage $E_B(\tau)$ comes at the cost of consuming this accumulated steering. In this sense, the steering can  be regarded as an essential consumable resource for energy storage. Fig. \ref{fig4}(b) focuses on illustrating the evolution of systems within reservoir characterized by high dissipation. It can be concluded that, firstly, the increase in dissipation significantly disrupts both the internal energy and steering within the system. Secondly, while heightened dissipation amplifies the asymmetry of EPR steering, it does not alter steering's influence on energy dynamics. That is, after breaking equilibrium, stored energy continues to grow at the expense of steering consumption until reaching its peak.}

\textcolor{black}{Figs. \ref{fig4}(c)-(d) reveal the variation of energy in reservoirs at non-zero temperatures. As shown in Fig. \ref{fig4}(c), the system's energy dissipates more rapidly into the reservoir, as evidenced by the faster decline in the energy peak compared to Fig. \ref{fig4}(a). However, this difference does not affect the relationship between steering and energy observed in Fig. \ref{fig4}(a). In the high-temperature fermionic reservoir depicted in Fig. \ref{fig4}(d), no significant energy dissipation was observed, yielding results nearly identical to those in Fig. \ref{fig4}(a). Interestingly, despite the complete transfer of energy, the party receiving all the energy was unable to steer the other. That is, $S_{B\to A}<0$ when $E_B=\omega_0$ or $S_{A\to B}<0$ when $E_A=\omega_0$. This outcome stands in stark contrast to the results obtained from the zero-temperature reservoir. }

 \textcolor{black}{\subsection{Resonance-driven Charging}}

\textcolor{black}{In this subsection, we investigate the impact of pump presence on system evolution, ensuring both the charger and battery have opportunities to simultaneously occupy excited states. Additionally, we conduct comparisons across different initial states  to uncover the relationship between steering and energy in details.}

\textcolor{black}{Figs. \ref{fig5}(a)-(c) show the time evolution of an initially fully empty charging system. First, regarding energy storage, pumpless charging demonstrates superiority over resonance-driven charging, as the former achieves approximately twice charging power $P(\tau)=E_B(\tau)/\tau$ of the latter. Second, it can be observed that the extremum of steering occurs at the system's energy equilibrium point $E_A=E_B$, and the steering peak will emerge at  $E_A=E_B=0.5\omega_0$. In addition, the steering prioritizes the energy balance of the charging system; even when both the charger and battery are excited, it cannot generate the steerability. In short, the steering accumulates as a resource during the initial stage of energy growth, serving as a resource to be consumed before energy reaches its peak. The phenomenon is identical to that observed in the pumpless charging scenario.}

\textcolor{black}{Figs. \ref{fig5}(d)-(f) illustrate the evolution of an initial high-energy charger and empty battery system. Compared to the initial fully depleted system in Figs. \ref{fig5}(a)-(c), the initial high-energy charger does not demonstrate the  advantage, as the battery's fully charged state persists for a shorter duration. Notably, the maximum steering no longer strictly occurs at $E_A=E_B=0.5\omega_0$, but rather at an energy equilibrium point near this location. However, this particular case does not undermine steering  as an essential resource for energy enhancement.}

 \textcolor{black}{\subsection{Steering and Battery}}

\textcolor{black}{Current research has revealed the essential relationship between steering and systemic energy: steering facilitates rapid balancing of systemic energy, and the sustained growth of stored energy requires steering as its fuel. In this subsection, we delve into the connection between steering and the battery itself, encompassing the battery's extractable work and its population transfer status.}

\textcolor{black}{Since ${\sigma _z} = \left| e \right\rangle \left\langle e \right| - \left| g \right\rangle \left\langle g \right|$, we have ${\left\langle {{\sigma _z}} \right\rangle _B} = {\rm Tr}\left[ {{\sigma _z}{\rho _B}\left( t \right)} \right] = {p_e} - {p_g}$, where $p_e$ and $p_g$ represent the populations of the excited state and ground state, respectively. In particular, when ${\left\langle {{\sigma _z}} \right\rangle _B} =  - 1$, it indicates that the battery's population is concentrated in the ground state, meaning the battery is completely depleted $E_B=0$. Conversely, the population has fully shifted to the excited state when ${\left\langle {{\sigma _z}} \right\rangle _B} =   1$, i.e., the battery is fully charged $E_B=\omega_0$. When ${\left\langle {{\sigma _z}} \right\rangle _B} =  0$, it suggests a balanced population distribution between the battery's high and low energy states, meaning the battery has reached half its capacity $E_B=0.5\omega_0$.}

\begin{figure}
		\begin{minipage}{0.5\textwidth}
			\centering
		\subfigure{\includegraphics[height=5cm]{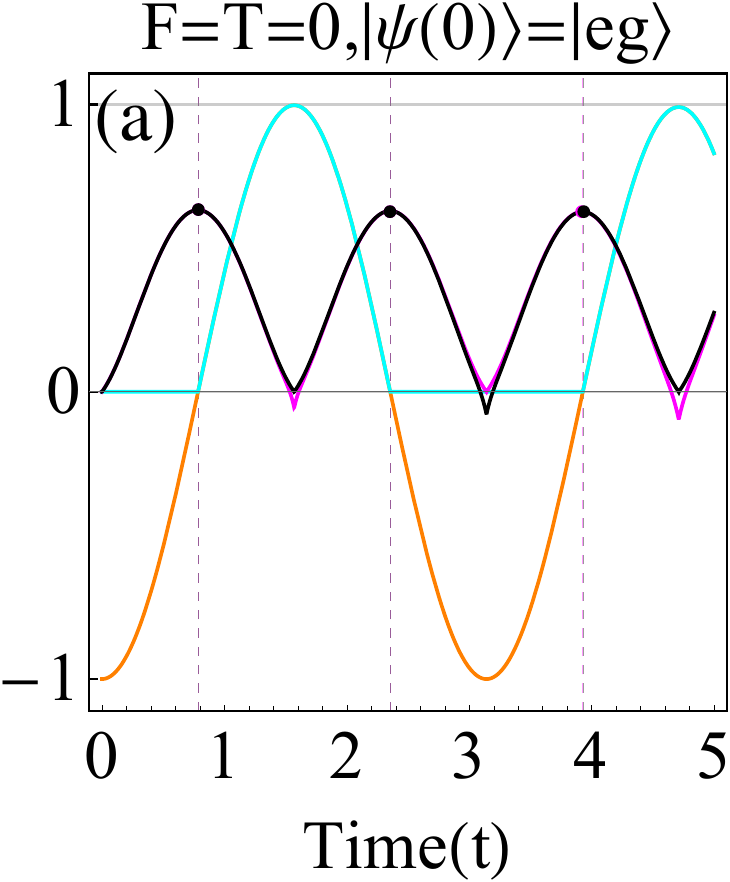}}
		\hspace{0.1cm}
		\subfigure{\includegraphics[height=5cm]{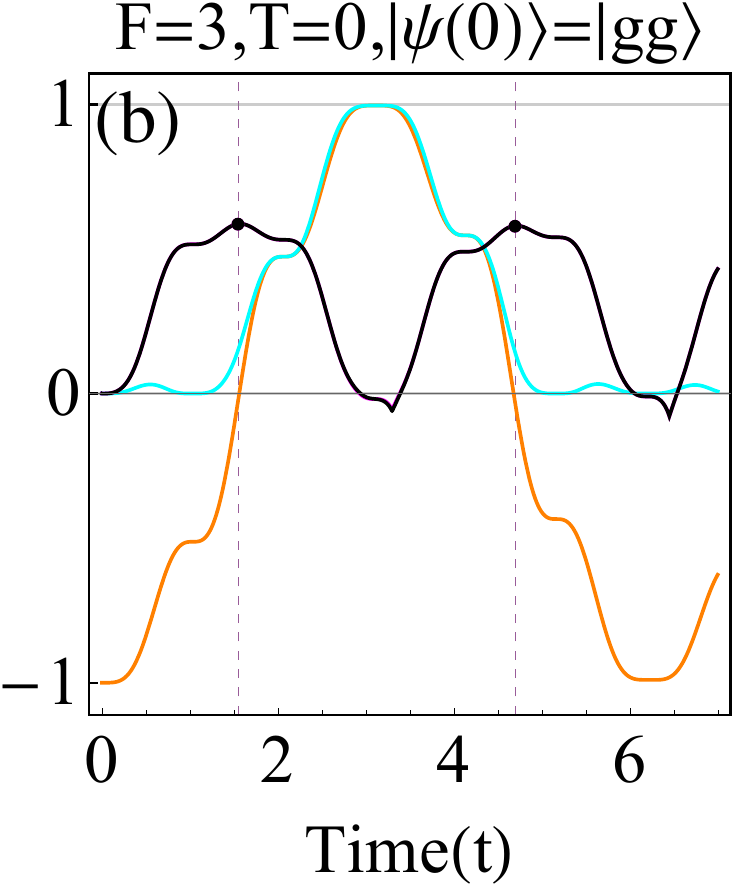}}
		\\
		\subfigure{\includegraphics[height=5cm]{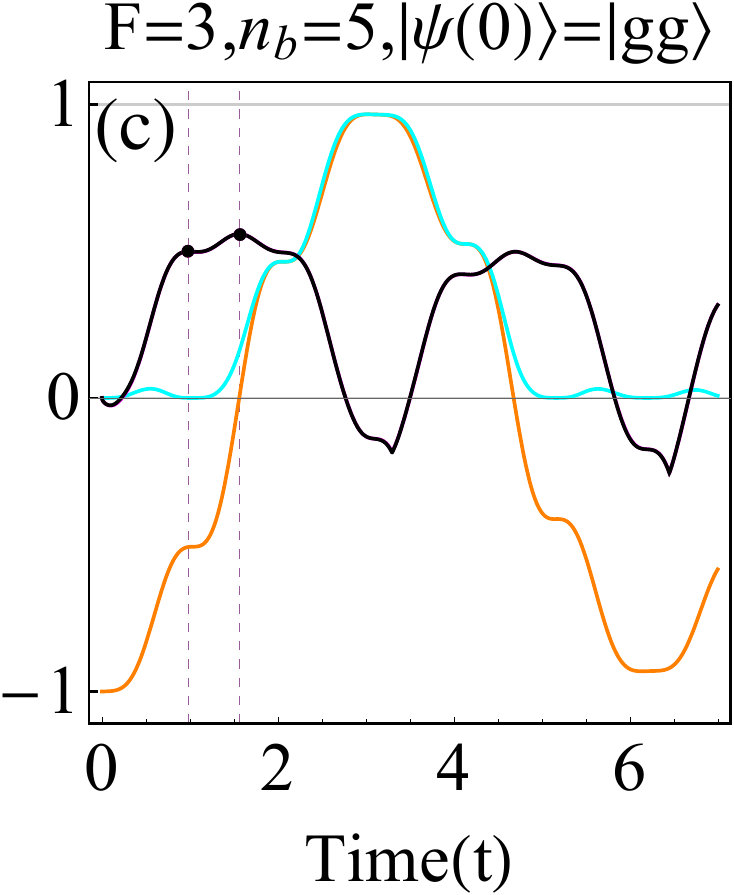}}
		\hspace{0.1cm}
		\subfigure{\includegraphics[height=5cm]{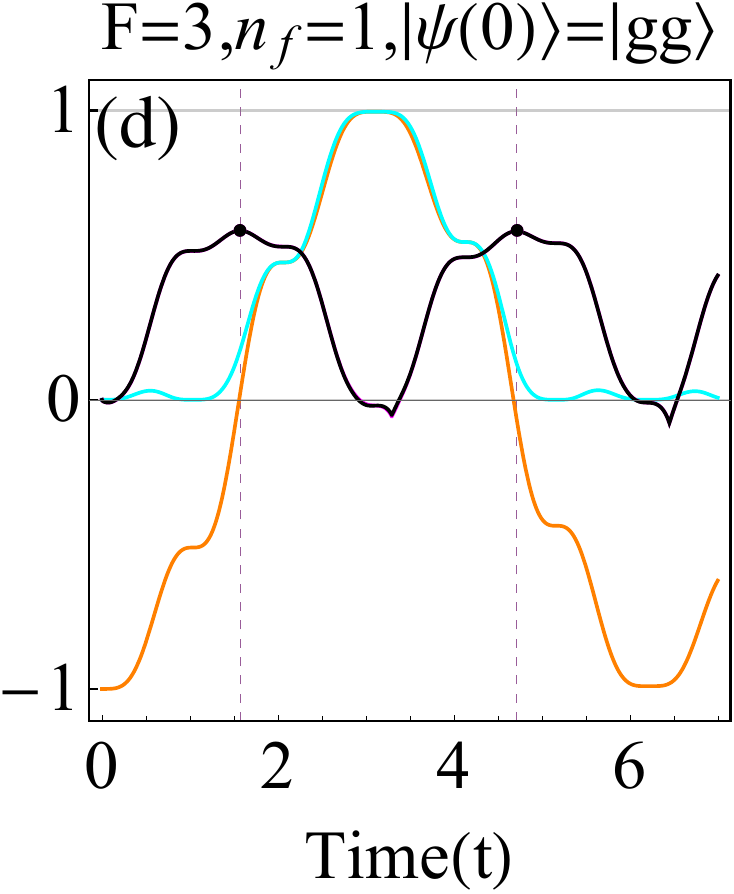}}
          \\
        \subfigure{\includegraphics[width=5cm]{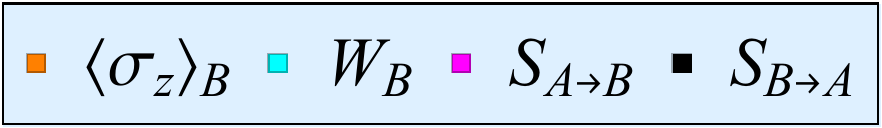}}
		\end{minipage}\hfill
		\caption{\textcolor{black}{Battery populations difference ${\left\langle {{\sigma _z}} \right\rangle _B}$, ergotropy $W_B$ and steering function $S_{A\to B, (B\to A)}$ as a function of time. We considered different charging scenarios and initial system conditions. Common to all figures are the resonance-driven and low-dissipation reservoirs, i.e., $\Delta=0,\Gamma=0.001,g=1$. Note that the magenta dots and black dots represent the maxima of steering functions $S_{A\to B}$ and $S_{B\to A}$, respectively, while the corresponding dashed lines indicate the time points at which these maxima occur.}}
		\label{fig6}
	\end{figure}

\textcolor{black}{As shown in the Fig. \ref{fig6}(a), the maximum steering occurs at the equilibrium point of the battery's population distribution, and the stored energy is about to be converted into ergotropy. The enhancement of ergotropy is essentially the process of consuming steering, which parallels the continuous improvement process of energy storage. The conclusion can be extended to resonance-driven scenarios and is independent of reservoir temperature, as detailed in Figs. \ref{fig6}(b)-(d). That is, steering always emphasizes the equilibrium of the battery's population, while the growth of ergotropy invariably comes at the expense of steering.}

 \section{Discussions and conclusions}

In summary, we investigate the dynamical properties of energy and quantum steering of charging systems in charging protocol, \textcolor{black}{where charger and battery dissipate energy into a shared reservoir. To be explicit, we study the relationship between EPR steering and the energy of the charging system from three perspectives while ensuring the system could obtain high energy. The first is to demonstrate the} \textcolor{black}{existence of optimal parameter configurations that enable the system to reach its maximum energy.} The results show that \textcolor{black}{strong drive, low dissipation, and high-temperature environments are essential  for realizing high-energy systems.}  Interestingly, \textcolor{black}{in high-temperature fermionic reservoirs, increasing the detuning between the system and the drive proves to be a crucial operation for enhancing energy.} Secondly, \textcolor{black}{we observe the relationship between steering and system energy under various charging scenarios, discovering that steering serves as an essential resource for enhancing energy storage, accumulating prior to the system's energy equilibrium to function as the consumable for achieving peak energy storage.} Third, we find that \textcolor{black}{optimizing steering equates to pursuing battery's population balance, so continuous improvements in energy storage and extractable work inevitably come at the expense of steering. It is worth noting that since our investigation is based on low-dissipation mechanisms, the relationship between steering and system energy similarly applies to the charger-battery model across different reservoir configurations.} Finally, \textcolor{black}{we supplement our investigation with high-dissipation charging in the \hyperlink{A}{Appendix A}. The results reveal that due to the absence of steerability, the complex relationship between steering potential and energy can only be discussed with appropriate caution. Given that the steering consistently provides stable indications of energy changes within complex charging scenarios, we believe it would serve as an excellent energy indicator in the future, accurately monitoring and predicting energy variations across more high-performance QBs protocols.}

\vskip 0.5cm
	
	\begin{acknowledgements}
	This work was supported by the National Science Foundation of
China (Grant nos. 12475009, 12075001, and 62471001), Anhui Provincial Key Research and Development Plan (Grant No. 2022b13020004),
Anhui Provincial Science and Technology Innovation Project (Grant No.
202423r06050004), Anhui Provincial University Scientific Research
Major Project (Grant No. 2024AH040008), Anhui Provincial Natural Science Foundation (Grant No. 202508140141),  and Anhui
Provincial Department of Industry and Information Technology (Grant No. JB24044).	
        
	\end{acknowledgements}

\appendix
\textcolor{black}{\hypertarget{A}{\section{Steering and Energy in High-Dissipation Regimes}}}

\textcolor{black}{The main text of the investigation has demonstrated the contribution of steering to energy growth. This result is predicated on our establishment of a low-dissipation environment to achieve faster and greater energy storage. In this appendix, we consider charging stability by enhancing reservoir dissipation, thereby further investigating the role of steering in energy.}

\begin{figure}
		\begin{minipage}{0.5\textwidth}
			\centering
		\subfigure{\includegraphics[height=4.5cm]{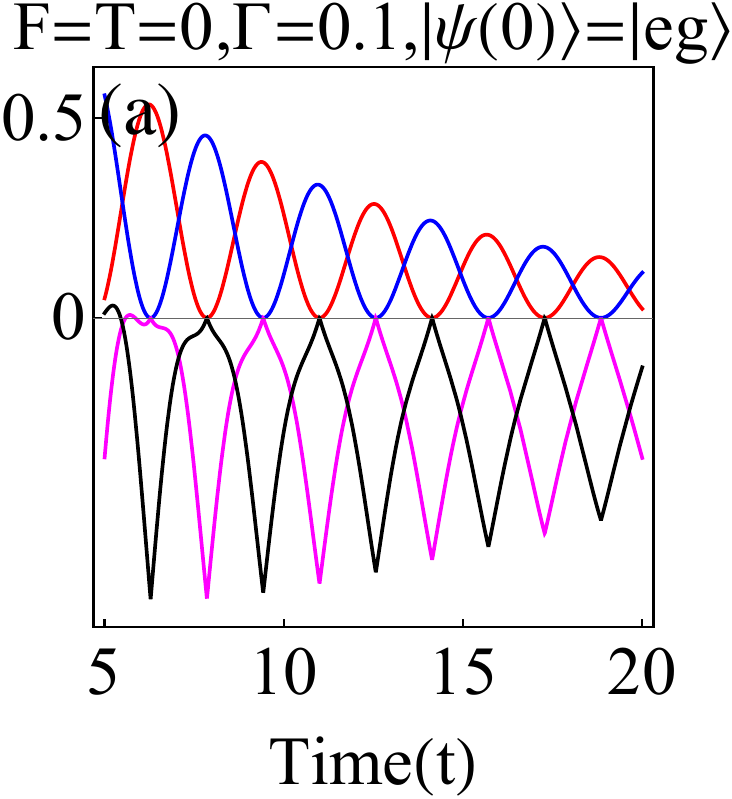}}
		\hspace{0.1cm}
		\subfigure{\includegraphics[height=4.5cm]{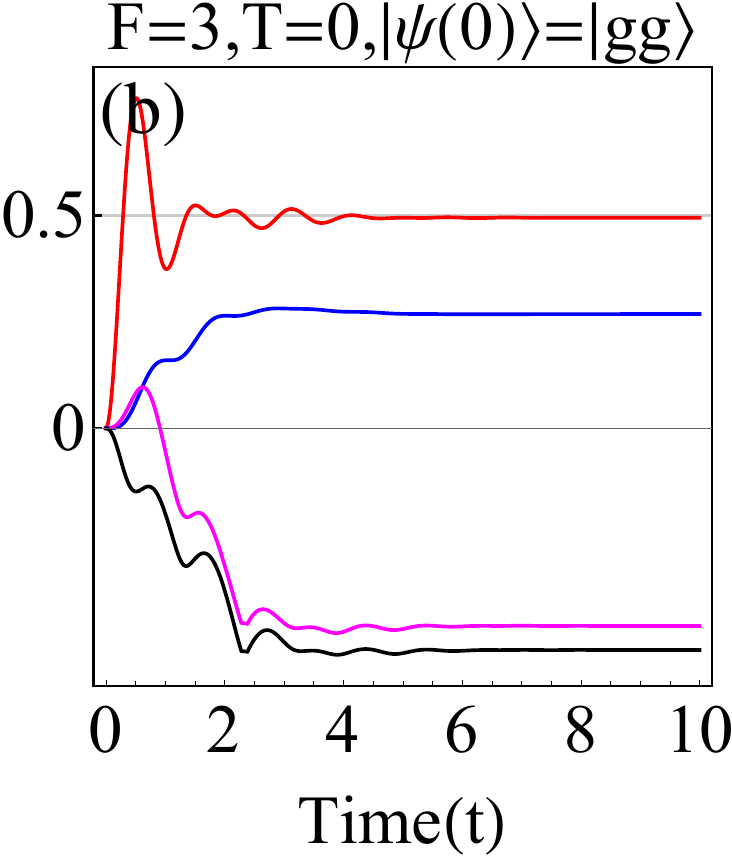}}
		\\
		\subfigure{\includegraphics[height=4.5cm]{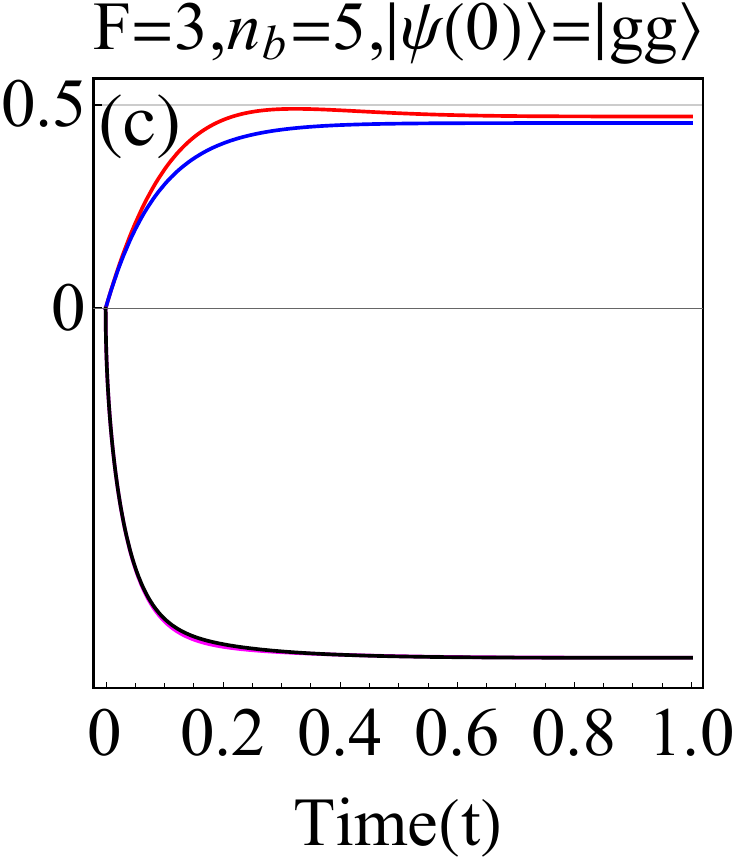}}
		\hspace{0.1cm}
		\subfigure{\includegraphics[height=4.5cm]{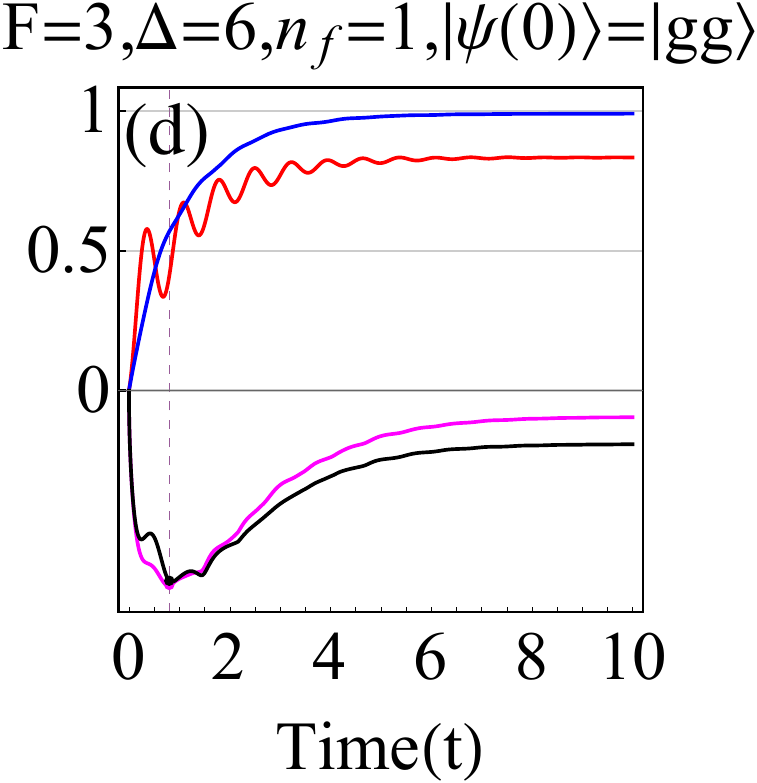}}
          \\
        \subfigure{\includegraphics[width=5cm]{457_lengend.pdf}}
		\end{minipage}\hfill
		\caption{\textcolor{black}{System energy and steering as a function of time. Multiple charging scenarios and high-dissipation reservoirs are considered here. The specific settings are as follows: Graph (a)-(c): $\Delta=0,g=1$. Graph (b)-(d): $\Gamma=1,g=1$.}}
		\label{fig7}
	\end{figure}

\textcolor{black}{Fig. \ref{fig7} displays different charging scenarios in high-dissipation reservoirs. Most notably, steerability is virtually absent—meaning the steering potential has always remained unactivated. Consequently, this investigation serves as an additional supplement to the relationship between steering and energy. It can be seen that during pumpless charging in Fig. \ref{fig7}(a), the steering potential exhibits a positive correlation with energy (i.e., ${E_A} \propto {S_{A \to B}}, {E_B} \propto {S_{B \to A}}$), meaning energy growth is always accompanied by improvements in the steering potential of the subsystem. With the emergence of resonance-driven charging (cf. Figs. \ref{fig7}(b)-(c)), the steering potential then transforms into a consumable resource for energy growth (i.e., ${E_B} \propto {S_{A \to B}^{-1}}, {S_{B \to A}^{-1}}$). Most unusually, in the fermionic reservoir (cf. Fig. \ref{fig7}(d)), energy enhancement first requires suppressing the steering potential, only to re-stimulate its growth once energy surpasses approximately half the capacity.}

\textcolor{black}{Therefore, for high-dissipation charging regimes, the contribution of steering to energy is concentrated during the stage when steering potential remains unactivated. The effect of steering potential on energy is complex and is dependent on the specific charging scenarios.}

\end{document}